\documentclass[conference]{IEEEtran}
\IEEEoverridecommandlockouts
\usepackage{cite}
\usepackage{comment}
\usepackage{amsmath,amssymb,amsfonts}
\usepackage{nomencl}
\usepackage{etoolbox}
\usepackage{algorithmic}
\usepackage{graphicx}
\usepackage{stfloats}
\usepackage{textcomp}
\usepackage{xcolor}
\usepackage{tikz}
\usetikzlibrary{positioning}
\usetikzlibrary{shapes.geometric, arrows}
\def\BibTeX{{\rm B\kern-.05em{\sc i\kern-.025em b}\kern-.08em
    T\kern-.1667em\lower.7ex\hbox{E}\kern-.125emX}}

\begin{document}

\tikzstyle{io} = [rectangle, rounded corners, minimum width=1.5cm, minimum height=1cm,  text centered, yshift=-0.5cm, draw=black, align=left]
\tikzstyle{block} = [rectangle, rounded corners, minimum width=1.5cm, minimum height=1cm,  text centered,  yshift=-0.5cm, draw=black, align=left]
\tikzstyle{process} = [rectangle, minimum width=1.5cm, minimum height=1cm, text  centered, yshift=-0.5cm, draw=black, align=center]
\tikzstyle{arrow} = [thick,->,>=stealth]

\title{Envisioning security control in renewable dominated power systems through stochastic multi-period AC security constrained optimal power flow 
\thanks{This research work has received funding from the European Union's Horizon 2020 research and innovation program under grant agreement No 864298 (project ATTEST).}
}

\author{\IEEEauthorblockN
{
Mohammad Iman Alizadeh, Muhammad Usman, Florin Capitanescu }
\IEEEauthorblockA{\textit{Environmental Research and Innovation (ERIN) Department, Luxembourg Institute of Science and Technology (LIST)} \\
Belvaux, Luxembourg \\
mohammad.alizadeh@list.lu, muhammad.usman@list.lu, florin.capitanescu@list.lu}

}

\maketitle

\begin{abstract}
The accelerated penetration rate of renewable energy sources (RES) brings environmental benefits at the expense of increasing operation cost and undermining the satisfaction of the N-1 security criterion. To address the latter issue, this paper envisions N-1 security control in RES dominated power systems through stochastic multi-period AC security constrained optimal power flow (SCOPF). The paper extends the state-of-the-art, i.e. deterministic and single time period AC SCOPF, to capture two new dimensions, RES stochasticity and multiple time periods, as well as emerging sources of flexibility such as flexible loads (FL) and energy storage systems (ESS). Accordingly, the paper proposes and solves for the first time a new problem formulation in the form of stochastic multi-period AC SCOPF (S-MP-SCOPF). The S-MP-SCOPF is formulated as a non-linear programming  (NLP) problem. It computes optimal setpoints of flexibility resources and other conventional control means for congestion management and voltage control in day-ahead operation. Another salient feature of this paper is the comprehensive and accurate modelling, using: AC power flow model for both pre-contingency and post-contingency states, inter-temporal constraints for resources such as FL and ESS in a 24-hours time horizon and RES uncertainties. The importance and performances of the proposed model through a direct approach, pushing the problem size up to the solver limit, are illustrated on two test systems of 5 nodes and 60 nodes, respectively, while future work will develop a tractable algorithm. 
\end{abstract}

\begin{IEEEkeywords}
congestion management, energy storage systems, flexibility, flexible loads, security-constrained optimal power flow, voltage control
\end{IEEEkeywords}

\renewcommand\nomgroup[1]{%
  \item[\bfseries
  \ifstrequal{#1}{A}{Sets}{%
  \ifstrequal{#1}{B}{Parameters}{%
  \ifstrequal{#1}{C}{Variables}{}}}%
]}
\makenomenclature
\nomenclature[A]{$N$}{Set of nodes indexed by $n$}
\nomenclature[A]{$S$}{Set of scenarios indexed by $s$}
\nomenclature[A]{$T$}{Set of time periods indexed by $t$}
\nomenclature[A]{$K$}{Set of operation states, including normal and contingencies, indexed by $k$}
\nomenclature[A]{$F$}{Set of FL indexed by $f$}
\nomenclature[A]{$E$}{Set of ESS indexed by $e$ }
\nomenclature[A]{$R$}{Subset of nodes with RES}
\nomenclature[A]{$G$}{Set of generators indexed by $g$}
\nomenclature[B]{$c_{g}$}{cost (€/MWh) of active power re-dispatch of\\ generator $g$}
\nomenclature[B]{$\Delta T$}{time interval ratio }
\nomenclature[B]{$\pi_{s}$}{probability of occurrence of scenario $s$}
\nomenclature[B]{$c_{e}$}{cost (€/MWh) of active power of energy storage $e$ }
\nomenclature[B]{$c_{f}$}{cost (€/MWh) of active power of flexible load $f$ }
\nomenclature[B]{$c_{n}^{Gcurt}$}{cost (€/MWh) of active power curtailment at node $n$ }
\nomenclature[B]{$c_{n}^{Lcurt}$}{cost (€/MWh) of active load curtailment at node $n$ }
\nomenclature[C]{$P_{G_{g,s,t}}^{k}$}{active power of generator $g$ at time $t$, scenario $s$, state $k$ }
\nomenclature[B]{$P_{G_{g,t}}^{*}$}{active power of generator $g$ at time $t$ cleared in the energy market }
\nomenclature[C]{$P_{e,s,t}^{ch,k}$}{active power charging of storage $e$ at time $t$, scenario $s$, state $k$ }
\nomenclature[C]{$P_{e,s,t}^{dis,k}$}{active power discharging of storage $e$ at time $t$, scenario $s$, state $k$ }
\nomenclature[C]{$P_{f,s,t}^{inc,k}$}{active power increase of FL $f$ at time $t$, scenario $s$, state $k$ }
\nomenclature[C]{$P_{f,s,t}^{dec,k}$}{active power decrease of FL $f$ at time $t$, scenario $s$, state $k$ }
\nomenclature[C]{$Rc_{n,s,t}^{k}$}{active power of RES curtailment at node $n$, time $t$, scenario $s$n state $k$ }
\nomenclature[C]{$Lc_{n,s,t}^{k}$}{active load curtailment at node $n$, time $t$, scenario $s$, state $k$ }
\nomenclature[C]{$Q_{c_{n,s,t}}$}{reactive load curtailment at node $n$, time $t$, scenario $s$, state $k$ }
\nomenclature[C]{$P_{inj_{n,s,t}}^{k}$}{active power injection at node $n$, time $t$, scenario $s$, state $k$ }
\nomenclature[B]{$P_{D_{n,t}}$}{load active power at node $n$, period $t$ }
\nomenclature[B]{$R_{n,s,t}$}{RES active power at node $n$, period $t$, scenario $s$ }
\nomenclature[C]{$Q_{G_{g,s,t}}^{k}$}{reactive power of generator $g$ at time $t$, scenario $s$, state $k$ }
\nomenclature[B]{$Q_{D_{n,t}}$}{load reactive power at node $n$, period $t$ }
\nomenclature[C]{$Q_{inj_{n,s,t}}^{k}$}{reactive power injection at node $n$, time $t$, scenario $s$, state $k$ }
\nomenclature[C]{$e_{n,s,t}^{k}$}{real part of complex voltage $\left(e_{n,s,t}^{k}+jf_{n,s,t}^{k}\right)$ at node $n$, period $t$, scenario $s$, state $k$ }
\nomenclature[C]{$f_{n,s,t}^{k}$}{imaginary part of complex voltage at node $n$, period $t$, scenario $s$, state $k$ }
\nomenclature[B]{$G_{nm}$}{conductance of the branch linking nodes $n$ and $m$}
\nomenclature[B]{$B_{nm}$}{susceptance of the branch linking nodes $n$ and $m$}
\nomenclature[B]{$B^{sh}_{nm}$}{shunt susceptance of the branch linking nodes $n$ and $m$}
\nomenclature[B]{$\underline{P}_{g}/\overline{P}_{g}$}{minimum/maximum active power limit of generator $g$}
\nomenclature[B]{$\underline{Q}_{g}/\overline{Q}_{g}$}{minimum/maximum reactive power limit of generator $g$}
\nomenclature[B]{$I_{nm}^{\max}$}{maximum current of line linking nodes $n$ and $m$}
\nomenclature[B]{$\underline{V}_{n}/\overline{V}_{n}$}{minimum/maximum voltage limit at node $n$}
\nomenclature[B]{$\Delta P_{G_{g}}$}{ramp rate limit of generator $g$}
\nomenclature[C]{$SOC_{e,s,t}^{k}$}{State-of-Charge for storage $e$ at period $t$, scenario $s$, state $k$ }
\nomenclature[B]{$SOC_{e}^{\min}$}{minimum State-of-Charge for storage $e$ }
\nomenclature[B]{$SOC_{e}^{\max}$}{maximum State-of-Charge for storage $e$ }
\nomenclature[B]{$\overline{P}_{e}^{ch}/\overline{P}_{e}^{dis}$}{maximum active power charging/discharging limit of storage $e$ }
\nomenclature[B]{$\overline{P}_{f}^{inc}/\overline{P}_{f}^{dec}$}{maximum active power increase/decrease limit of FL $f$ }
\nomenclature[B]{$\eta^{dis,e}$}{discharging efficiency rate of ESS $e$ }
\nomenclature[B]{$\eta^{ch,e}$}{charging efficiency rate of ESS $e$ }
\printnomenclature

\section{Introduction}

\subsection{Motivation}

To attain the stringent sustainable goals set to them, power systems worldwide are hosting increasingly large amounts of renewable energy sources (RES) at all voltage levels. However, massive RES penetration significantly challenges the enforcement of transmission system security \cite{balu1992line} due to the inherent variability and difficulty to predict RES output. 
In this context, power systems operate closer to their security limits and hence fulfilling N-1 security becomes a challenging task, particularly under stressed operation conditions, unexpected RES output, and/or unavailability of effective control actions. Regarding the latter aspect, as classical control means (e.g. conventional power plants) and controllable RES may not be sufficient to fulfill security, additional emerging sources of flexibility, such as flexible loads (FL) and energy storage systems (ESS), are being deployed to enhance power system flexibility and offset issues provoked by RES \cite{alizadeh2016flexibility,wang2016enhancing}. 

Deterministic AC security-constrained optimal power flow (SCOPF) \cite{capitanescu2011state,stott2012optimal,capitanescu2018challenges} is the conventional tool to enforce N-1 security at a given period of time. SCOPF is mainly used in the day-ahead operation for the cost-optimal procurement of ancillary services (e.g. for managing congestion and voltages). To this end, SCOPF computes the optimal balance of preventive (i.e. pre-contingency) and corrective (i.e. post-contingency) actions able to guarantee static system security (i.e. pertaining to congestion and voltage magnitude) for a set of postulated (e.g. N-1) contingencies. 

A variety of SCOPF problems have been tackled, that target optimizing the redispatch of either active or reactive powers \cite{capitanescu2006applications,capitanescu2011state}. SCOPF problems are in their simplest form formulated as large scale (non-convex) nonlinear programs (NLPs), whose main difficulty is the large size \cite{capitanescu2011state,stott2012optimal,capitanescu2018challenges}. 

\subsection{Related Works}

Deterministic single period AC SCOPF is state-of-the-art \cite{phan2013some,platbrood2013generic,jiang2013novel,kardovs2019two, avramidis2020novel,de2018security, madani2015promises,9252174}. Its solution has been extensively explored through various algorithms: decomposition methods (e.g. Benders decomposition or iterative algorithms based on contingencies filtering, both embedding interior-point method for core NLP problem) applied to exact formulations \cite{phan2013some,platbrood2013generic,jiang2013novel,kardovs2019two}, approximations \cite{platbrood2013generic,avramidis2020novel}, meta-heuristics \cite{de2018security}, and even convex relaxations (e.g. semi-definite programming \cite{phan2013some,madani2015promises} and second order cone programming \cite{9252174}) that are able to assess the optimality gap of exact algorithms' solution. Further modeling advancement regarding generators' response after contingencies to frequency and voltage control have been also explored \cite{avramidis2020novel,valencia2021fast}. 

Solving deterministic single period AC SCOPF is today computationally demanding but still scalable to systems of reasonably large size (i.e. thousand nodes) \cite{capitanescu2018challenges}. Despite AC SCOPF is state-of-the-art, some works develop sophisticated algorithms for its linear (DC) SCOPF approximation via column and constraint generation \cite{velloso2021exact}, constraints redundancy screening \cite{9094021}, alternating direction method of multipliers (ADMM) in a distributed manner \cite{velay2019fully}, network compression \cite{karbalaei2018new}, or machine learning \cite{velloso2021combining}.


To capture RES inherent variability, two timely extensions of SCOPF have been developed independently to address: 
\begin{itemize}
\item {\it uncertainties} (regarding RES) based on robust optimization \cite{capitanescu2012cautious}, distributionally robust optimization \cite{you2021risk}, stochastic optimization (exact \cite{vrakopoulou2013probabilistic}, simplified DC \cite{8810833} or relaxed \cite{venzke2018convex}), and chance-constrained optimization \cite{hamon2013value}; other uncertainties (e.g. regarding corrective control potential failure) were tackled via chance-constraints \cite{karangelos2019iterative}.  
\item {\it multiple time periods} (linked by inter-temporal constraints) via the simplified DC model \cite{sharifzadeh2017multi,murillo2013secure} or the exact AC model \cite{fuchs2017security,schanen2018toward}. 
\end{itemize}
However, these extensions are very scarce and tremendously increase the computational burden of the problem. 

Additionally, to reliably deal with RES variability, a meaningful SCOPF problem should also consider time dependent emerging flexibility resources (e.g. FL and ESS). However, these flexibility resources were considered only sporadically and in a single period deterministic SCOPF \cite{cao2015improved}.

\subsection{Paper Contributions and Organization}

It can be concluded that the approaches aimed to extend SCOPF state-of-the-art are not only very scarce but also have considered separately the two main features: RES uncertainties and multiple time periods. In addition, the approaches to any of these two challenges do not model the two other difficult features as AC network model and emerging flexibility resources in a joint fashion. 

To bridge this gap, as a conceptual contribution, this paper proposes the new envisioned concept of multi-period AC SCOPF under uncertainties to control N-1 security in RES-dominated power systems of the future. The main contribution of this paper is the extension of the state-of-the-art, i.e. deterministic AC SCOPF, to capture jointly two new dimensions (RES stochasticity and multiple time periods) as well as the emerging sources of flexibility (FL and ESS). In other words, the paper proposes and solves for the first time a new problem formulation in the form of a stochastic multi-period AC SCOPF (S-MP-SCOPF). 

A direct approach relying on the state-of-the-art NLP solver IPOPT \cite{wachter2002ipopt}, widely used in many AC OPF/SCOPF applications, is conducted formulating the largest problem size that the solver can still manage while a tractable algorithm is planned for future work. 
Note that, the ``size challenge'' of the proposed S-MP-SCOPF problem is determined by the product of four different dimensions: the size of the system, number of postulated contingencies, number of uncertainty scenarios, and number of time periods.

To further highlight the above mentioned novel contributions of this work, Table \ref{compare} summarizes the main modelling features of the proposed approach which distinguishes it from the several existing methods. One can observe that, like this work, scalability is not pursued per se in most works that address more challenging AS SCOPF problem extensions. Also, it is implied that if AC grid model is not used, then simplified models (e.g. DC) are adopted. 

\setlength{\tabcolsep}{1.5pt}
\begin{table}[htb]
\centering 
\caption{Model features of various approaches}
\label{compare}
\begin{tabular}{ccccccc} \hline
model & deterministic & multiple    & operation  & flexibility & AC  & scala- \\
      & single-period & time periods& uncertainty& resources   & model   & bility             \\\hline
\cite{phan2013some}-\cite{velloso2021exact} &\text{\sffamily x}& &  &  & \text{\sffamily x}   &\text{\sffamily x}\\
\cite{9094021},\cite{velloso2021combining} & \text{\sffamily x}& & & & & \text{\sffamily x} \\
\cite{velay2019fully}& \text{\sffamily x}& & & & & \\
\cite{capitanescu2012cautious}&&  & \text{\sffamily x} &  & \text{\sffamily x}\\
\cite{you2021risk}&&  & \text{\sffamily x} &  & & \text{\sffamily x} \\
\cite{vrakopoulou2013probabilistic} &  & \text{\sffamily x} &  & \text{\sffamily x}\\
\cite{8810833} &&  & \text{\sffamily x} &  & & \text{\sffamily x}\\
\cite{venzke2018convex} &&  & \text{\sffamily x} &  & \text{\sffamily x}\\
\cite{hamon2013value} &&  & \text{\sffamily x} &  & \text{\sffamily x}\\
\cite{karangelos2019iterative}& &  & \text{\sffamily x} &  & \text{\sffamily x}\\
\cite{sharifzadeh2017multi}&& \text{\sffamily x} & \text{\sffamily x} &  & & \text{\sffamily x}\\
\cite{murillo2013secure}& & \text{\sffamily x} & \text{\sffamily x} & \text{\sffamily x} & & \text{\sffamily x}\\
\cite{fuchs2017security}& & \text{\sffamily x} &  &  & \\
\cite{schanen2018toward}& & \text{\sffamily x} &  &  & \text{\sffamily x}\\
proposed && \text{\sffamily x} & \text{\sffamily x} & \text{\sffamily x} & \text{\sffamily x}\\ \hline 
\end{tabular}
\end{table}
\setlength{\tabcolsep}{6pt}

The remaining of the paper is organized as follows. Section II presents the detailed formulation of the S-MP-SCOPF problem. 
Section III provides quantitative results with a direct approach to the proposed problem. Section IV concludes and 
provide directions for future works. 

\section{Formulation of the S-MP-SCOPF Problem}

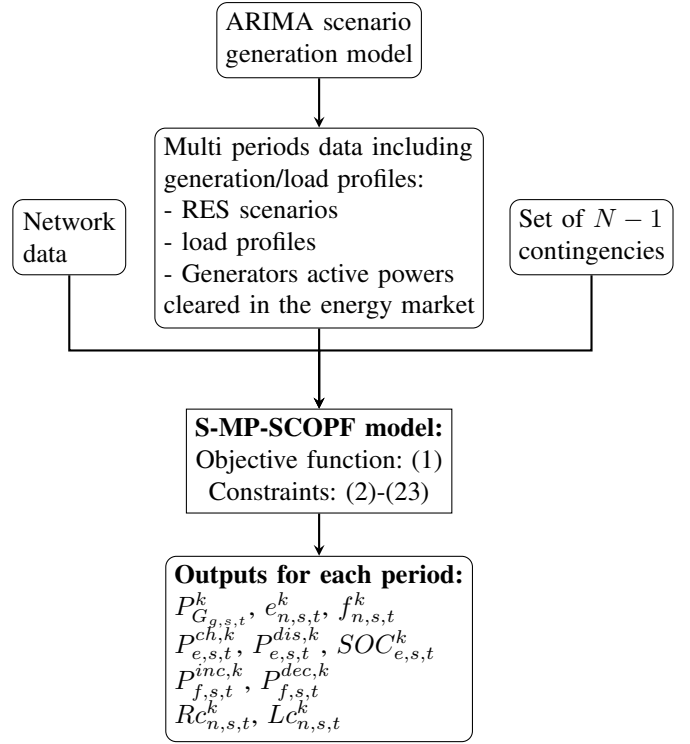
\begin{figure}[htb]
    \centering
\begin{tikzpicture}[node distance=2cm]
\node (in1) [block] {ARIMA scenario \\generation model}
;
\node (in2) [io, below of=in1] { 
Multi periods data including\\ generation/load profiles:\\ 

- RES scenarios\\
- load profiles\\
- Generators active powers \\
   cleared in the energy market
};
\node (in3) [io, below left of=in1, xshift=-1.2cm, node distance=3cm] { 
Network\\ data

};\node (in4) [io, below right of=in1, xshift=1.5cm, node distance=3cm] { 

Set of $N-1$\\ contingencies

};
\node (pro1) [process, below of=in2,yshift=-0.6cm,] {\textbf{S-MP-SCOPF model:}\\
Objective function: (\ref{eq1})\\
Constraints: (\ref{eq2})-(\ref{eq23})
};
\node (out1) [block, below of=pro1] {\textbf{Outputs for each period:}\\
$P_{G_{g,s,t}}^{k}$, $e_{n,s,t}^{k}$, $f_{n,s,t}^{k}$\\
$P_{e,s,t}^{ch,k}$, $P_{e,s,t}^{dis,k}$, $SOC_{e,s,t}^{k}$\\
$P_{f,s,t}^{inc,k}$, $P_{f,s,t}^{dec,k}$\\
$Rc_{n,s,t}^{k}$, $Lc_{n,s,t}^{k}$
};
\draw [arrow] (in1) -- (in2);
\draw [arrow] (in3.south) -- ++(0,-1cm) -|  (pro1);
\draw [arrow] (in4.south) -- ++(0,-1cm) -|  (pro1);
\draw [arrow] (in2.south) -- ++(0,-1cm) -|  (pro1);
\draw [arrow] (pro1) -- (out1);
\end{tikzpicture}
 \caption{A flowchart of the proposed model.}
    \label{flowchart}
\end{figure}

This section describes in detail the proposed S-MP-SCOPF model to procure, in day-ahead operation planning, flexibility for congestion management and voltage control such that to satisfy N-1 security criterion. The model computes optimal setpoints for flexibility resources (FL and ESS), RES curtailment, and other conventional control means (e.g. generators) in each time period and system state, as illustrated in Fig. \ref{flowchart}. The model relies on AC power flow equations expressed using voltages in rectangular coordinates. 
\par
The objective \eqref{eq1} of the S-MP-SCOPF is to minimize the expected cost of flexibility procurement for ancillary services (congestion and voltage control) in transmission network operation under both normal and post contingency states. This cost pertains to the re-dispatch of conventional generators, ESS, and FL, curtailment of RES, and load curtailment to prevent infeasibility.    

\begin{align} \label{eq1}
&\min \sum_{s\in S}\sum_{t \in T} \pi_{s}
\Bigg\{ 
\sum_{g\in G} \left(P_{G_{g,s,t}}^{0}-P_{G_{g,t}}^{*}\right)c_{g} \nonumber\\
&+\sum_{ k\in K} \Bigg [ \sum_{e\in E}
\left(P_{e,s,t}^{ch,k}+P_{e,s,t}^{dis,k}\right)c_{e}\nonumber\\
&+\sum_{f \in F}  \left(P_{f,s,t}^{inc,k}+P_{f,s,t}^{dec,k}\right)c_{f}\nonumber\\
&+\sum_{n\in R}
\left(Rc_{n,s,t}^{k}\right)c_{n}^{Gcurt}
+\sum_{n\in N}
\left(Lc_{n,s,t}^{k}\right)c_{n}^{Lcurt}
\Bigg ]
\Bigg\}
\end{align}
The problem is subject to the following constraints:
\begin{align} 
&\sum_{g\in G} P_{G_{g,s,t}}^{k} + R_{n,s,t}
+\sum_{e\in E}\left(P_{e,s,t}^{dis,k}-P_{e,s,t}^{ch,k}\right)\nonumber\\
&+\sum_{f \in F}\left(P_{f,s,t}^{dec,k}-P_{f,s,t}^{inc,k}\right)
-Rc_{n,s,t}^{k}+Lc_{n,s,t}^{k}=\nonumber\\
&P_{D_{n,t}}+ P_{inj_{n,s,t}}^{k} \nonumber \\
&\forall{n \in N, s \in S, t \in T, k \in K}\label{eq2}
\end{align}
\begin{align}
&\sum_{g\in G} Q_{G_{g,s,t}}^{k}=
Q_{D_{n,t}}-Q_{c_{n,s,t}}+ Q_{inj_{n,s,t}}^{k} \nonumber \\
&\forall{n \in N, s \in S, t \in T, k \in K}\label{eq3}
\\
&P_{inj_{n,s,t}}^{k}=\left[\left(e_{n,s,t}^{k}\right)^2+
\left(f_{n,s,t}^{k}\right)^2\right]\sum_{m \in N}G_{nm}\nonumber\\
&-\sum_{m \in N}\big[\left(e_{n,s,t}^{k}e_{m,s,t}^{k}
+f_{n,s,t}^{k}f_{m,s,t}^{k}\right)G_{nm}\nonumber \\
&+\left(f_{n,s,t}^{k}e_{m,s,t}^{k}-e_{n,s,t}^{k}f_{m,s,t}^{k}
\right)B_{nm}
\big] \nonumber\\
&\forall{n \in N, s \in S, t \in T, k \in K}\label{eq4}
\\
&Q_{inj_{n,s,t}}^{k}=-\left[\left(e_{n,s,t}^{k}\right)^2+
\left(f_{n,s,t}^{k}\right)^2\right]
\sum_{m \in N}\left(B_{nm}^{sh}+B_{nm} \right)\nonumber\\
&+\sum_{m \in N}\big[\left(e_{n,s,t}^{k}e_{m,s,t}^{k}
+f_{n,s,t}^{k}f_{m,s,t}^{k}\right)B_{nm}\nonumber\\
&-\left(f_{n,s,t}^{k}e_{m,s,t}^{k}-e_{n,s,t}^{k}f_{m,s,t}^{k}
\right)G_{nm}
\big]\nonumber\\
&\forall{n \in N, s \in S, t \in T, k \in K}\label{eq5}\\
&\underline{P}_{g}\leq P_{G_{g,s,t}}^{k}\leq\overline{P}_{g}
\hspace{2em} \forall{g \in G, s \in S, t \in T, k \in K }
\label{eq6}\\
&\underline{Q}_{g}\leq Q_{G_{g,s,t}}^{k}\leq\overline{Q}_{g}
\hspace{2em} \forall{g \in G, s \in S, t \in T, k \in K }
\label{eq7}\\
&\left(G_{nm}^2+B_{nm}^2\right)\big[\left(e_{n,s,t}^{k}-e_{m,s,t}^{k}\right)^{2}
+
\left(f_{n,s,t}^{k}-f_{m,s,t}^{k}\right)^{2}
\big]\nonumber\\
&\leq\left(I_{nm}^{\max}\right)^2
\hspace{4.5em} \forall{n,m \in N, s \in S, t \in T, k \in K }
\label{eq8}
\\
&\left(\underline{V}_{n}\right)^2\leq
\left(e_{n,s,t}^{k}\right)^{2}
+
\left(f_{n,s,t}^{k}\right)^{2}\leq
\left(\overline{V}_{n}\right)^2
\nonumber\\
&\hspace{8em} \forall{n \in N, s \in S, t \in T, k \in K }
\label{eq9}\\
&   \left|P_{G_{g,s,t-1}}^{0}-P_{G_{g,s,t}}^{0}\right|\leq \Delta P_{G_{g}} \hspace{1em}  \forall{g \in G, s \in S, t \in T }
\label{eq10}\\
  & \left|P_{G_{g,s,t}}^{k}-P_{G_{g,s,t}}^{0}\right|\leq \Delta P_{G_{g}}\nonumber\\
  & \hspace{7em} \forall{g \in G, s \in S, t \in T, k \in K,k\neq0 }
\label{eq11}\\
&  SOC_{e,s,t+1}^{k}=SOC_{e,s,t}^{k}+
    \Delta T\left(\eta^{ch,e} P_{e,s,t}^{ch,k}-P_{e,s,t}^{dis,k}/\eta^{dis,e}\right)
    \nonumber\\
& \forall{e \in E, s \in S, t \in T, k \in K }   \label{eq12}
\\
& SOC_{e}^{min}\leq SOC_{e,s,t}^{k}\leq SOC_{e}^{max}
    \nonumber\\
& \forall{e \in E, s \in S, t \in T, k \in K }   \label{eq13} \\
& SOC_{e,s,T}^{k}=SOC_{e,s,0}^{k}, \forall{e \in E, s \in S, t \in T, k \in K } \label{eq14} \\
& 0\leq P_{e,s,t}^{ch,k}\leq \overline{P}_{e}^{ch}, \forall{e \in E, s \in S, t \in T, k \in K }   \label{eq16} \\
& 0\leq P_{e,s,t}^{dis,k}\leq \overline{P}_{e}^{dis}, \forall{e \in E, s \in S, t \in T, k \in K }   \label{eq17}\\ 
&  \frac{P_{e,s,t}^{ch,k}}{\overline{P}_{e}^{ch}} + \frac{P_{e,s,t}^{dis,k}}{\overline{P}_{e}^{dis}} \leq 1, \forall{e \in E, s \in S, t \in T, k \in K } \label{eq15} \\
& \sum_{t \in T}  P_{f,s,t}^{inc,k}   =   \sum_{t \in T}  P_{f,s,t}^{dec,k}, \forall{f\in F, s \in S, t \in T, k \in K }  \label{eq18} 
\end{align}
\begin{align}
& 0\leq   P_{f,s,t}^{inc,k}\leq \overline{P}_{f}^{inc}, \forall{f \in F, s \in S, t \in T, k \in K }   \label{eq19} \\
& 0\leq   P_{f,s,t}^{dec,k}\leq \overline{P}_{f}^{dec}, \forall{f \in F, s \in S, t \in T, k \in K }  \label{eq20} \\
& \frac{P_{f,s,t}^{dec,k}}{ \overline{P}_{f}^{dec}} + \frac{P_{f,s,t}^{inc,k}} {\overline{P}_{f}^{inc}} \leq 1, \forall{f \in F, s \in S, t \in T, k \in K }  \label{eq21} \\
& 0 \leq   Lc_{n,s,t}^{k}\leq \ P_{D_{n,t}}, \forall{n \in N, s \in S, t \in T, k \in K }    \label{eq22} \\
& 0 \leq   Rc_{n,s,t}^{k}\leq \ R_{n,s,t}, \forall{n \in N, s \in S, t \in T, k \in K }    \label{eq23} 
\end{align}
where $k=0$ represents normal operation state while $k\geq 1$ indicates contingency states, all notations being defined in the nomenclature. 

Constraints \eqref{eq2} and \eqref{eq3} represent active and reactive power balance equations (for each node $n$, scenario $s$, time $t$ and state $k$), which include active/reactive power flows from Eqs \eqref{eq4} and \eqref{eq5}. Note that in \eqref{eq3} load curtailment assumes constant power factor. 

Constraints \eqref{eq6} and \eqref{eq7} are the hard physical limits on active and reactive powers of generator $g$. 

Network operation constraints (congestion and voltages) are modeled by constraints \eqref{eq8} and \eqref{eq9}. Eq \eqref{eq8} represents the longitudinal branch current limit, which is a reasonable approximation of the current aimed to avoid doubling the number of constraints (e.g. when the current is expressed for both ends of the branch). Eq \eqref{eq9} imposes limits on node voltage magnitude. 

Eq \eqref{eq10} restricts the ramping of generator $g$ for each two successive time intervals of normal operating state. 
Eq \eqref{eq11} is the coupling constraint on active power of generator $g$ between normal operation and post-contingency states. 

The ESS operation is captured by the following set of constraints \cite{gayme2012optimal}. Eq \eqref{eq12} describes the dynamics of State-of-Charge (SoC), \eqref{eq13} is the SoC limit for each ESS, \eqref{eq14} maintains the the SoC of ESS equal on first and last time periods, Eqs \eqref{eq16} and \eqref{eq17} are limits on active power charging and discharging of ESS period, and \eqref{eq15} prevents simultaneous charging and discharging of storage $e$ for each period. 

It is important to note that \eqref{eq15} is a smart and tractable exact approximation, proposed in \cite{shen2020modeling} to avoid introducing binary variables to model the statuses charging and discharging of an ESS. In the latter work it is demonstrated that this modeling matches accurately the effect of using binary variables, i.e. at the optimum an ESS either charges or discharges but not both. This is due to the fact that both charging and discharging statuses have associated costs in the objective function which in turn prevents the simultaneous charging and discharging of an ESS. This effect is also empirically observed in all our numerical simulations. 

The FL operation is modeled by the following set of constraints. Eq \eqref{eq18} maintains the energy balance of a FL over whole time horizon, \eqref{eq19} and \eqref{eq20} are the limits on the increase and decrease of active power of FL, respectively, and Eq \eqref{eq21} prevents simultaneous increase and decrease in the active power of FL during each time interval. Remark that \eqref{eq21} relies on the same type of assumption and approximation as for storage elements in Eq \eqref{eq15}. 

Finally, \eqref{eq22} limits the load curtailment while \eqref{eq23} restricts the RES curtailment. 

Note that the proposed S-MP-SCOPF is an NLP problem. 

Last but not least, a modelling aspect worth discussing for any stochastic optimization problem is the number of decision-making stages assumed and interpretation/implementation of the optimal stochastic solution. Often, in day-ahead operation planning, there are two such stages corresponding to ``here and now'' decisions and ``wait and see'' decisions. While two stages can be straightforwardly modeled, we opted only for modelling all decisions as ``wait and see'' (i.e. scenario-dependent, as opposed to ``here and now'' decisions which are scenario independent) for the sake of computational challenge, as it leads to a larger stochastic optimisation challenge. In such case, the transmission system operator can either implement in practice the optimal solution corresponding to one of the assumed scenarios or a weighted (e.g. via probabilities of occurrence) of solutions of all postulated scenarios.

\section{Numerical Results}

The features of the proposed S-MP-SCOPF model are illustrated using two test systems of 5 and 60 nodes respectively, for 24-hours time frame (one hour time resolution), given sets of N-1 contingencies and different number of scenarios. 

All simulations are performed in Julia/JuMP open source programming language \cite{dunning2017jump} on a PC of 2.11 GHz and 48 GB of RAM. IPOPT optimizer is used to solve all NLP problems \cite{wachter2002ipopt}. 

\subsection{Results for 5-node test system}

\begin{figure}
  \begin{center}
    \includegraphics[scale=0.55]{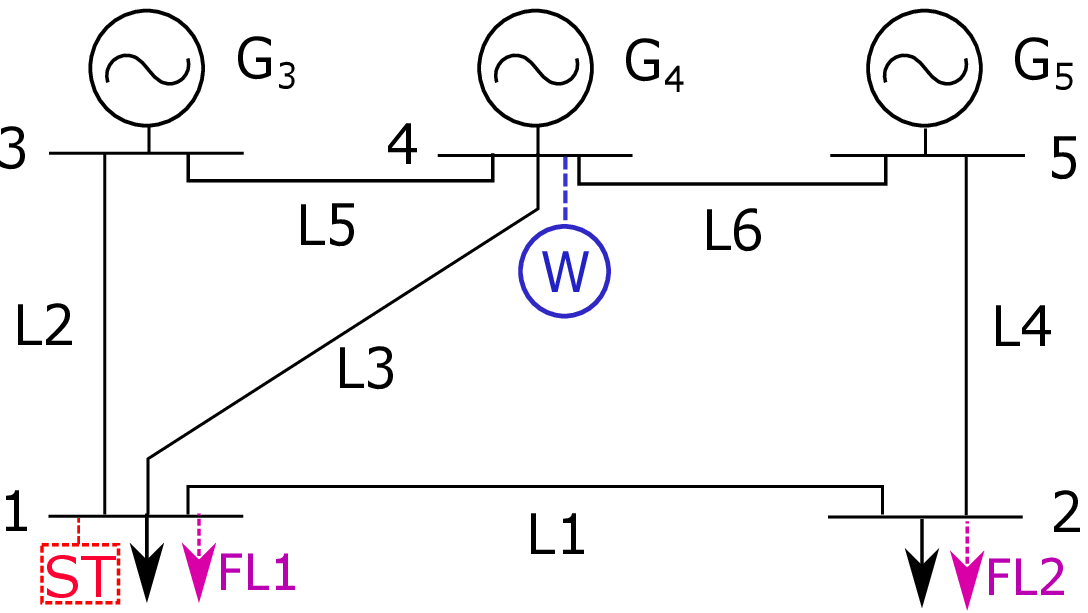}
  \end{center}

  \caption{\small One-line diagram of the 5-bus system}
  \label{diagram}
\end{figure}

The 5 node system is adapted from \cite{capitanescu2018challenges} and its one-line diagram is shown in Fig. \ref{diagram}. Tables \ref{data1} and \ref{data_branch} respectively show the steady-state and line data for the 5-node test system \cite{capitanescu2018challenges}. We consider 6 N-1 line contingencies and up to 10 uncertainty scenarios. Full results for this test case are comprehensively discussed and all necessary data are provided to enable benchmarking, reproducibility and comparison.

\begin{table*}[htb]
\centering 
\caption{5-bus system data at peak load with no wind, no storage and no flexible load}
\label{data1}
\begin{tabular}{cccccccccccccccc} \hline
    & $P_L$ & $Q_L$ & $P_G$ & $Q_G$ & $V$   &$V^{\min}$ & $V^{\max}$ & $P_G^{\min}$ & $P_G^{\max}$ & $Q_G^{\min}$ & $Q_G^{\max}$ & $\Delta P_G$& $a$ & $b$ & $c$  \\ 
bus &  MW   & MVar &  MW   & MVar  & pu    &  pu    & pu      &    MW     & MW        & MVar      & MVar  &    MW   & €/MWh$^2$ & €/MWh & €\\  \hline 
1   & 1100  &  400 &   -   &   -   & 0.954 &  0.92  &  1.05   &    -      &      -    &    -      &   -   &    -   & -& - & -  \\  
2   &  500  &  200 &   -   &   -   & 0.950 &  0.92  &  1.05   &    -      &      -    &    -      &   -   &    -   & -& - & -  \\  
3   &   -   &  -   & 700.0 &  69.5 &  1.0  &  0.92  &  1.05   &   150     &     1500  &  -500     &  750  &   200 & 0.01 & 25 & 100\\  
4   &   -   &  -   & 600.0 & 304.9 &  1.0  &  0.92  &  1.05   &   150     &     1500  &  -500     &  750  &   200 & 0.01 & 60 & 100\\  
5   &   -   &  -   & 333.8 & 146.9 &  1.0  &  0.92  &  1.05   &   150     &     1500  &  -500     &  750  &   200 & 0.01 & 30 &  100 \\ \hline 
\multicolumn{16}{l}{$a$,$b$,$c$: Cost coefficients of conventional generators in nonlinear form $aP_{G}^2+bP_{G}+c$}
\end{tabular}
\end{table*}


\begin{table}[htb]
\centering 
\caption{5-bus system: line data.}
\label{data_branch}
\begin{tabular}{ccccccccc} \hline
     & bus   & bus  &  $V^{nom}$ & $R_{nm}$  &  $X_{nm}$ & $B_{nm}$ & $I_{nm}^{\max}$     \\ 
line & $n$   & $m$  & kV        & $\Omega$ & $\Omega$ & $\mu$S &   A    \\  \hline 
L1   &  1    &  2   & 400       &  3.2   & 16     &  160  &  1587.7     \\  
L2   &  1    &  3   & 400       &  6.4   & 32     &  320  &  1587.7     \\  
L3   &  1    &  4   & 400       &  3.2   & 16     &  160  &  1587.7      \\     
L4   &  2    &  5   & 400       &  6.4   & 32     &  320  &  1587.7     \\   
L5   &  3    &  4   & 400       &  6.4   & 32     &  320  &  1587.7     \\   
L6   &  4    &  5   & 400       &  6.4   & 32     &  320  &  1587.7     \\ \hline 
\end{tabular}
\end{table}

To consider RES, a wind farm is deployed at node 4. Table \ref{wind_scenarios} shows ten normalized scenario profiles over 24-hour period which are generated using a time series based Auto regressive integrated moving average (ARIMA) model \cite{sharma2013wind}. The scenarios and contingencies are equiprobable and (for simplicity) loads are assumed constant for the entire 24 hour horizon. 

\begin{table}[htb]
\centering 
\caption{Wind power scenarios profile (normalized values)}
\label{wind_scenarios}
\resizebox{\columnwidth}{!}{%
\begin{tabular}{ccccccccccc} \hline
 $t$    & s1  & s2  & s3 & s4 & s5 & s6 & s7 & s8 & s9 & s10 \\ \hline
1&	0.17&	0.36&	0.15&	0.59&	0.11&	0.01&	0.37&	0.06&	0.17&	0.16 \\
2&	0.21&	0.53&	0.30&	0.65&	0.14&	0.03&	0.46&	0.16&	0.00&	0.26 \\
3&	0.07&	0.26&	0.24&	0.59&	0.26&	0.09&	0.09&	0.12&	0.02&	0.11 \\
4&	0.00&	0.26&	0.23&	0.39&	0.16&	0.15&	0.12&	0.15&	0.04&	0.00\\
5&	0.02&	0.30&	0.19&	0.25&	0.15&	0.13&	0.24&	0.40&	0.08&	0.01\\
6&	0.00&	0.19&	0.30&	0.06&	0.08&	0.17&	0.29&	0.58&	0.08&	0.01\\
7&	0.00&	0.11&	0.25&	0.04&	0.01&	0.46&	0.26&	0.73&	0.03&	0.00\\
8&	0.00&	0.01&	0.33&	0.16&	0.03&	0.33&	0.27&	0.88&	0.04&	0.01\\
9&	0.10&	0.01&	0.22&	0.25&	0.00&	0.27&	0.41&	0.90&	0.05&	0.15\\
10&	0.26&	0.04&	0.33&	0.10&	0.06&	0.09&	0.33&	0.39&	0.20&	0.16\\
11&	0.19&	0.07&	0.25&	0.14&	0.14&	0.11&	0.57&	0.40&	0.46&	0.13\\
12&	0.08&	0.15&	0.07&	0.23&	0.27&	0.15&	0.28&	0.96&	0.24&	0.04\\
13&	0.21&	0.08&	0.06&	0.22&	0.26&	0.27&	0.03&	0.88&	0.17&	0.04\\
14&	0.40&	0.15&	0.14&	0.13&	0.20&	0.29&	0.07&	0.89&	0.03&	0.16\\
15&	0.30&	0.04&	0.41&	0.10&	0.29&	0.11&	0.09&	0.53&	0.01&	0.09\\
16&	0.04&	0.03&	0.52&	0.14&	0.12&	0.04&	0.17&	0.15&	0.01&	0.05\\
17&	0.18&	0.02&	0.45&	0.15&	0.06&	0.17&	0.30&	0.03&	0.06&	0.11\\
18&	0.21&	0.03&	0.36&	0.10&	0.01&	0.12&	0.26&	0.00&	0.08&	0.29\\
19&	0.09&	0.08&	0.11&	0.01&	0.01&	0.04&	0.07&	0.07&	0.09&	0.16\\
20&	0.23&	0.09&	0.04&	0.02&	0.04&	0.03&	0.02&	0.06&	0.49&	0.18\\
21&	0.28&	0.03&	0.01&	0.10&	0.01&	0.03&	0.02&	0.03&	0.30&	0.04\\
22&	0.38&	0.00&	0.00&	0.34&	0.07&	0.04&	0.11&	0.05&	0.31&	0.09\\
23&	0.11&	0.00&	0.00&	0.37&	0.02&	0.10&	0.15&	0.04&	0.14&	0.41\\
24&	0.10&	0.02&	0.04&	0.27&	0.01&	0.06&	0.48&	0.13&	0.10&	0.89\\
\hline
\end{tabular}
}
\end{table}

Four case studies are developed to assess the capability of the proposed model (in all cases load and RES generation curtailment is allowed to prevent infeasible problems):
\begin{itemize}
	\item Case\#0: no FL or ESS are considered;  
	\item Case\#1: one ESS is embedded at node 1 with the parameters provided in Table \ref{ESS_data} and $c_{e}$ cost is set to 80 €/MWh; 
	\item Case\#2: 10\% of load at node 1 and 2 (FL1 and FL2 in Fig. \ref{diagram}) is assumed flexible and the $c_{f}$ cost is set to 80 and 40 €/MWh, respectively in all operation states;  
	\item Case\#3: both ESS and FL are allowed to take part in optimization, with the costs given above. 
\end{itemize}

\begin{table}
\centering 
\caption{ESS characteristics}
\label{ESS_data}
\begin{tabular}{ccccccccc} \hline
     & $SOC_{e}^{\min}$   & $SOC_{e}^{\max}$  &  $\overline{P}_{e}^{ch}$ & $\overline{P}_{e}^{dis}$  &  $\eta^{ch,e}$ & $\eta^{dis,e}$     \\ 
bus & MWh   & MWh      & MW & MW    \\  \hline 
1   &  660  &  2200    & 50 &  50  & 0.95  & 0.95     \\ \hline 
\end{tabular}
\end{table}

\subsubsection{Case\#0}

Table \ref{case0_result} compares the results of the proposed model for different RES capacities, where RC0-RC10, CG, LC stand for RES capacity (between 0 and 1,000 MW), conventional generation and load curtailment respectively. It can be observed that, as the penetration rate of RES increases, CG cost reduces gradually since the RES production is paid by feed-in-tariff. However, the cost of curtailed energy increases up to 105,294 € in RC10. This suggests that efficient utilization of flexibility resources can potentially reduce the amount of curtailed energy.   

\subsubsection{Case\#1}

Table \ref{case1_result} provides the proposed model results with ESS at node 1 for different RES capacities. In comparison with the base case (i.e. Case\#0) the curtailment cost is reduced up to 46\% (i.e. $(105,294-56,754)/105,294$€) and the total cost reduces by $1,631,997-1,595,907=36,090$€ for RC10. In addition, the flexibility added by ESS prevents load curtailment in case RC7.

\subsubsection{Case\#2}

Similar benefits are observed using FL in both nodes 1 and 2 as shown in Table \ref{case2_result}. Using the flexibility provided by FL causes no energy curtailment for RC7 and RC8. Even in the case RC10, the total curtailment cost is reduced to 62\% (i.e. $(105,294-40,001)/105,294$€). In addition, the total cost for RC10 is reduced by 2.75\% with respect to the base case.

\subsubsection{Case\#3}

The results for the proposed S-MP-SCOPF model considering both ESS and FL are summarized in Table \ref{case3_result}. It can be seen that no energy curtailment occurs for RC7 and RC9. In addition, the curtailment cost decreases by 90.5\% (i.e. $(105,294-10,001)/105,294$€). One can also observe that the total cost is reduced by 4\% with respect to the base case. 

Another important remark is the synergy benefit of simultaneously using FL and ESS flexibility sources as can be noticed by the reduced cost of FL (9.88\%) and ESS (27.2\%) in comparison to the results reported for the Case\#1 and Case\#2 in Tables \ref{case1_result} and \ref{case2_result}, respectively. 

The computation time of the NLP problem is short in the range of few tens of seconds. Despite the small system size (5 nodes), the corresponding S-MP-SCOPF problem includes 7 operation states, 24 time periods and 10 scenarios is roughly equivalent in size to solving an AC OPF problem for a system of cca. 8,400 nodes. 
 
\begin{table}[htb]
\centering 
\caption{Case\#0 SCOPF results for different RES capacities}
\label{case0_result}
\begin{tabular}{ccccccccc} \hline
\textbf{RES}& \textbf{RES}    & \textbf{CG}   & \textbf{LC }  & \textbf{Total }   & \textbf{Time}   \\
\textbf{Cases}& \textbf{(MW)} & \textbf{ cost (€)} & \textbf{cost (€)}  & \textbf{cost (€)} & \textbf{ (s)}    \\  \hline 
RC0&0&	1,693,208&	0&	1,693,208&	12\\
RC1&100&	1,676,410&	0&	1,676,410&    11\\
RC2&200&	1,659,782&	0&	1,659,782&	12\\
RC3&300&	1,643,324&	0&	1,643,324&	11\\
RC4&400&	1,627,036&	0&	1,627,036&	12\\
RC5&500&	1,610,917&	0&	1,610,917&	13\\
RC6&600&	1,594,967&	0&	1,594,967&	14\\
RC7&700&	1,578,949&	2,287&	1,581,236&	21\\
RC8&800&	1,560,524&	2,8060&	1,588,584&	25\\
RC9&900&	1,543,530&	57,444&	1,600,974&	25\\
RC10&1,000&	1,526,703&	105,294&	1,631,997&	26\\
 \hline 
\end{tabular}
\end{table}

\begin{table}[htb] 
\centering 
\caption{Case\#1 SCOPF results with ESS for different RES capacities}
\label{case1_result}
\resizebox{\columnwidth}{!}{%
\begin{tabular}{ccccccccc} \hline
 \textbf{RES}    & \textbf{CG Cost }   & \textbf{LC cost}  & \textbf{ESS cost} & \textbf{Total cost}   & \textbf{Time}   \\ 
\textbf{Cases} & \textbf{ (€)} & \textbf{(€)} & \textbf{(€)}  & \textbf{(€)} & \textbf{ (s)}    \\  \hline 
RC0&	1,693,208&	0&	0&	1,693,208&	12\\
RC1&	1,676,410&	0&	0&	1,676,410&	13\\
RC2&	1,659,782&	0&	0&	1,659,782&	13\\
RC3&	1,643,324&	0&	0&	1,643,324&	13\\
RC4&	1,627,036&	0&	0&	1,627,036&	13\\
RC5&	1,610,917&	0&	0&	1,610,917&	14\\
RC6&	1,594,967&	0&	0&	1,594,967&	16\\
RC7&	1,579,193&	0&	389&	1,579,582&	23\\
RC8&	1,563,451&	2,064&	4,390&	1,569,905&	27\\
RC9&	1,546,819&	26,052&	5,300&	1,578,171&	32\\
RC10&	1,530,821&	56,754&	8,332&	1,595,907&	27\\
 \hline 

\end{tabular}
}
\end{table}

\begin{table}[htb]
\centering 
\caption{Case\#2 SCOPF results with FL for different RES capacities}
\label{case2_result}
\resizebox{\columnwidth}{!}{%
\begin{tabular}{ccccccccc} \hline
 \textbf{RES}    & \textbf{CG Cost }   & \textbf{LC cost}  & \textbf{FL cost} & \textbf{Total cost}   & \textbf{Time}   \\ 
\textbf{Cases} & \textbf{ (€)} & \textbf{(€)} & \textbf{(€)}  & \textbf{(€)} & \textbf{ (s)}    \\  \hline 
RC0&	1,693,208&	0&	0&	1,693,208&	12\\
RC1&	1,676,410&	0&	0&	1,676,410&	13\\
RC2&	1,659,782&	0&	0&	1,659,782&	13\\
RC3&	1,643,324&	0&	0&	1,643,324&	13\\
RC4&	1,627,036&	0&	0&	1,627,036&	13\\
RC5&	1,610,917&	0&	0&	1,610,917&	13\\
RC6&	1,594,967&	0&	0&	1,594,967&	14\\
RC7&	1,579,165&	0&	550&	1,579,715&	26\\
RC8&	1,563,507&	0&	6,710&	1,570,217&	20\\
RC9&	1,548,656&	11,780&	10,899&	1,571,335&	29\\
RC10&	1,532,739&	40,001&	14,377&	1,587,117&	33\\
 \hline 
\end{tabular}
}
\end{table}

\begin{table}[htb]
\centering 
\caption{Case\#3 SCOPF results with ESS and FL for 10 RES capacities}
\label{case3_result}
\resizebox{\columnwidth}{!}{%
\begin{tabular}{ccccccccc} \hline
 \textbf{RES}    & \textbf{CG  }   & \textbf{LC }  & \textbf{FL }& \textbf{ESS } & \textbf{Total }  & \textbf{Time} \\ 
\textbf{Cases} & \textbf{ Cost (€)} & \textbf{Cost (€)} & \textbf{Cost (€)}  & \textbf{Cost (€)}  & \textbf{Cost (€)} &\textbf{(s)}  \\  \hline 
RC0&	1,693,208&	0&	0&	0&	1,693,208&14\\
RC1&	1,676,410&	0&	0&	0&	1,676,410&14\\
RC2&	1,659,782&	0&	0&	0&	1,659,782&14\\
RC3&	1,643,324&	0&	0&	0&	1,643,324&14\\
RC4&	1,627,036&	0&	0&	0&	1,627,036&14\\
RC5&	1,610,917&	0&	0&	0&	1,610,917&14\\
RC6&	1,594,967&	0&	0&	0&	1,594,967&15\\
RC7&	1,579,193&	0&	0&	389&	1,579,582&26\\
RC8&	1,563,707&	0&	491&	4,390&	1,568,588&24\\
RC9&	1,550,080&	0&	6200&	5,299&	1,561,579&23\\
RC10&	1,536,615&	10,001&	12,956&	6,065&	1,565,637&30\\
 \hline 
\end{tabular}
}
\end{table}

\begin{figure}[htb]
	\centering
	\includegraphics[width=0.90\linewidth]{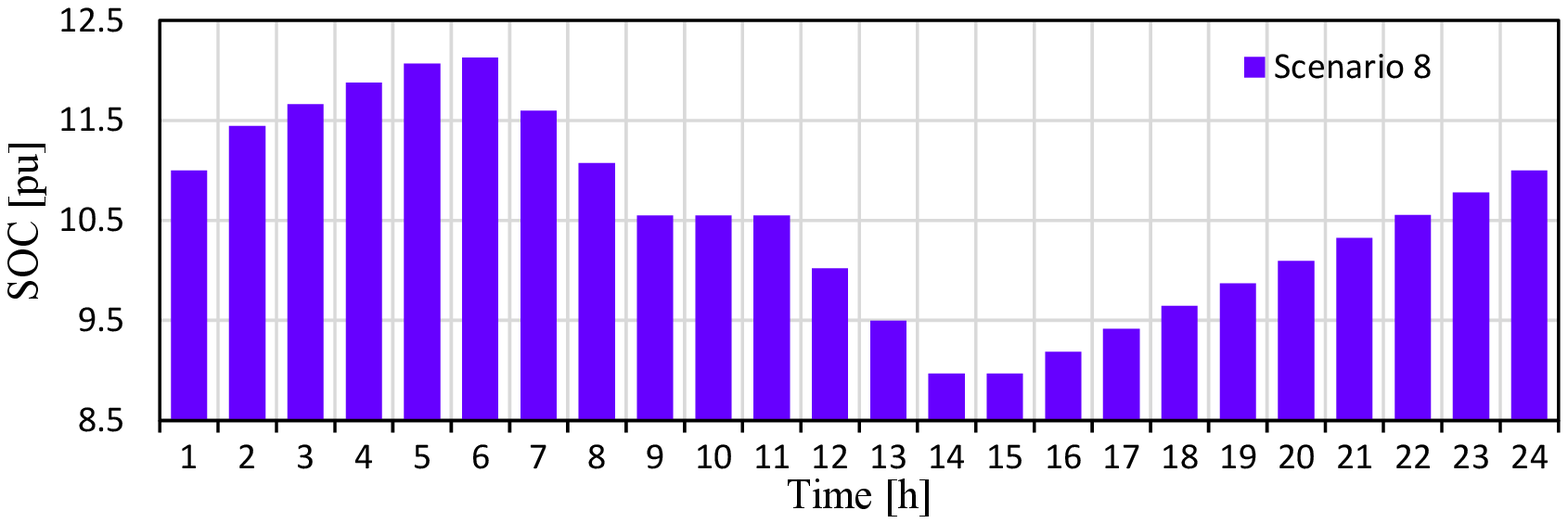}
	\caption{SOC profile of ESS for scenario 8 in contingency in line L2}
	\label{soc_5}
\end{figure}

\begin{figure}[htb]
	\centering
	\includegraphics[width=0.90\linewidth]{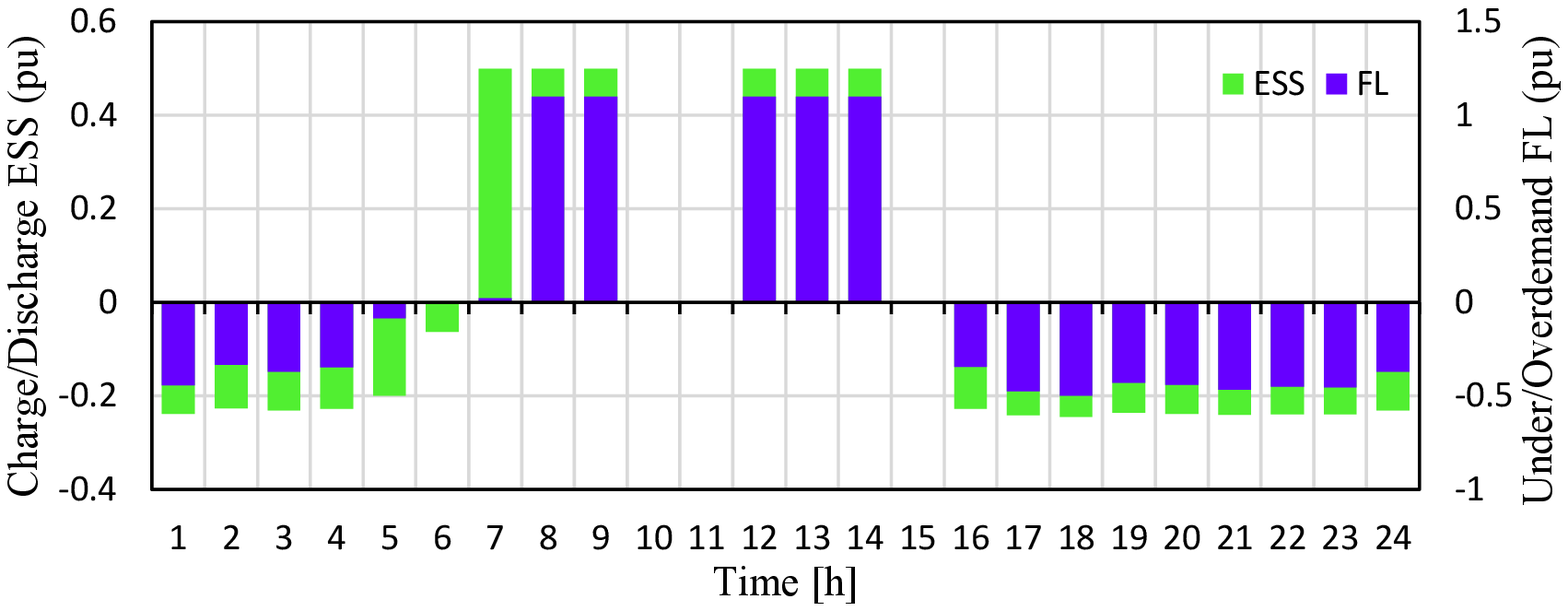}
	\caption{Injection and absorption profile of active power for both ESS and FL in contingency in line L2}
	\label{ess_fl_scen_8}
\end{figure}

For all case studies, the only binding contingency is the disconnection of line L2. For this contingency, in Case\#3 and RC10, Fig. \ref{soc_5} illustrates the state of charge (SoC) profile for scenario 8 (i.e. s8 in Table \ref{wind_scenarios}) while Fig. \ref{ess_fl_scen_8} plots and ESS and FL profiles. As expected, to accommodate maximum wind power in the network, ESS discharges in periods with excess of wind (i.e. 7, 8, 9, 12, 13 and 14) while, with the same trend, FL decreases the load (i.e. underdemand) during the same periods. To maintain their daily energy balance equal to zero, both ESS and FL increase charging and load to hours of low wind (i.e. 1-6 and 16-24).

\begin{figure}[htb]
	\centering
	\includegraphics[width=0.90\linewidth]{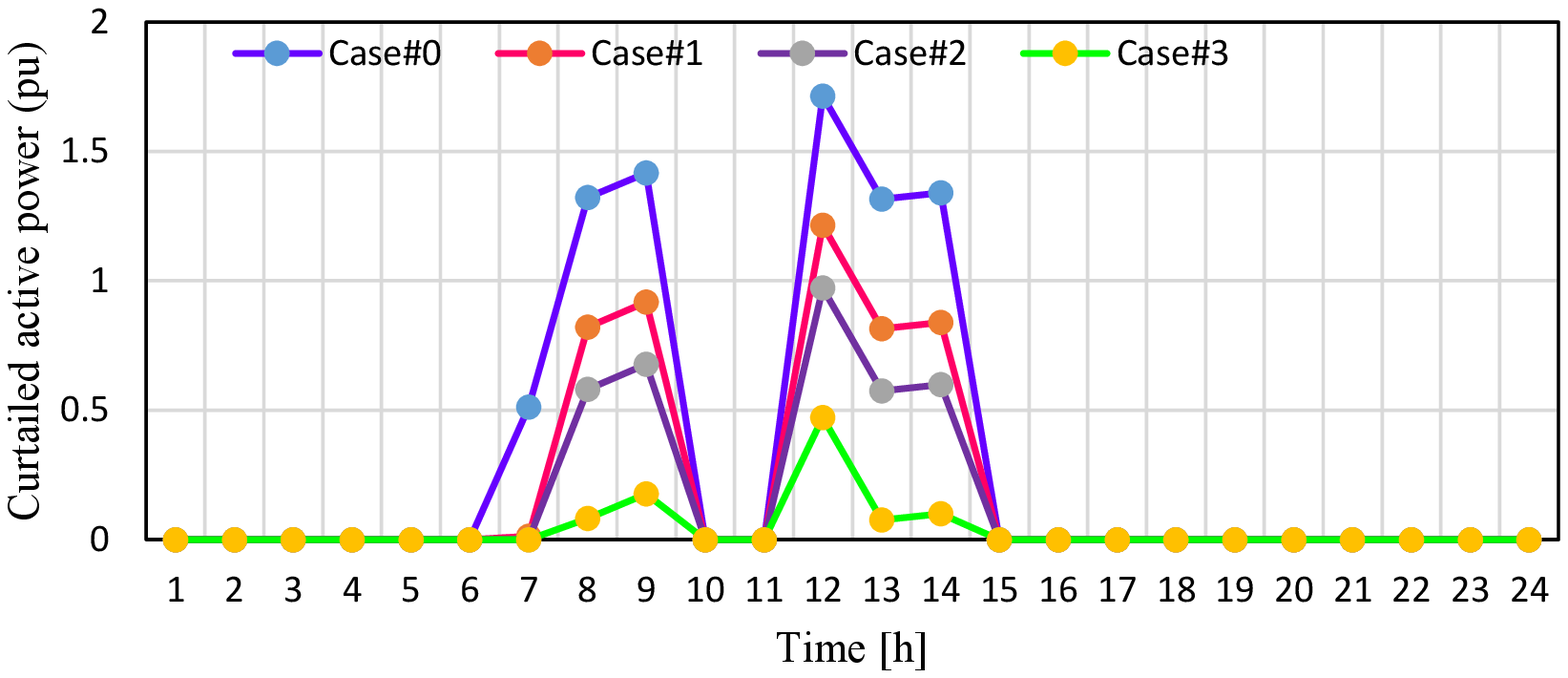}
	\caption{Curtailed active power profile for four cases: contingency L2 and s8}
	\label{curtailed_active}
\end{figure}

Fig. \ref{curtailed_active} illustrates the load curtailment for scenario s8 and contingency in line L2. A decreasing trend can be observed from Case\#0 to Case\#3 by considering the flexibility of ESS and FL. For instance, at 12 pm, the curtailed power reduces by 72\% (i.e. $(1.714-0.42)/1.714$) when considering both ESS and FL. This further demonstrates the benefits of additional flexibility offered by ESS and FL.

\subsection{Results for Nordic32 Test System}

\begin{figure}
	\centering
	\includegraphics[width=0.90\linewidth]{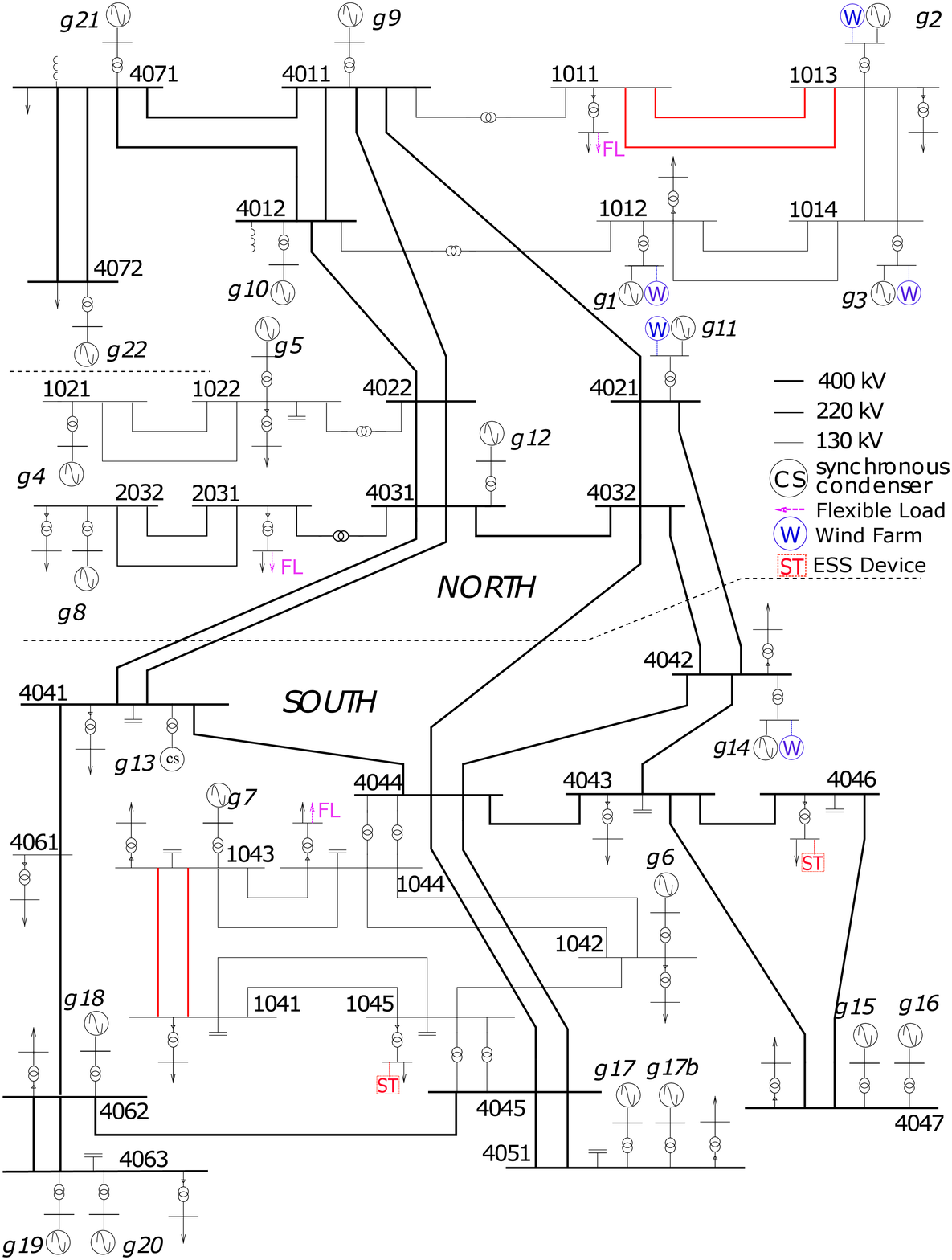}
	\caption{One-line diagram of Nordic32 test system}
	\label{nordic32}
\end{figure}

To test model scalability we use the synthetic Nordic32 test system \cite{van2013description}, which is closely inspired by the Sweden system. The test system includes 60 nodes, 23 generators, 57 lines, 31 transformers, and 12 shunts reactors/capacitors \cite{van2013description}. A contingency set of 33 N-1 line disconnections is assumed. We assume a futuristic renewable-dominated version of this system (see Fig. \ref{nordic32}), in which five large wind farms (with 7,200, 5,400, 6,300, 5,700, and 6,300 MW of rated power) are installed at nodes 1012, 1013, 1014, 4021 and 4042, respectively. As a consequence, to cope with the uncertain variability and potential excess of active power injected in north area while managing congestion and voltage issues, three FL are assumed at nodes 1011, 1044, and 2031, and two ESS (with the same parameters as in Table \ref{ESS_data}) are embedded at nodes 1045 and 4046. The load pattern from \cite{pena2017extended} is adopted for a generic summer day. To test the scalability of the proposed model, we consider up to 30 scenarios. 

\subsubsection{Illustration of flexibility resources}

To evaluate the added value of using emerging flexible resources (FL and ESS), like for the 5-bus system, four different case studies are defined, namely: Case\#0 (base case) in which neither FL nor ESS are used, Case\#1 where only ESS units are utilized, Case\#2 where only FL are considered, and finally, Case\#3 where both ESS and FL units are available. For all cases, 10 wind scenarios are generated, the ESS cost is set to 3 €/MWh and FL cost is set to 2.5€/MWh for both normal operation and post-contingency states. The load and generation curtailment cost is set to 30€/MWh i.e. ten times larger than the most expensive conventional generator cost. 

\begin{table*}[htb]
\centering 
\caption{Nordic32 test system results for different case studies}
\label{60bus_cases}
\begin{tabular}{ccccccccccc} \hline
&\multicolumn{5}{c|}{normal operation state}&\multicolumn{4}{|c|}{post contingency state}\\
\cline{2-11} 
 \textbf{Cases}    & \textbf{CG}   & \textbf{LC}  & \textbf{GC}& \textbf{FL} & \textbf{ESS}   & \textbf{LC}  & \textbf{GC}& \textbf{FL} & \textbf{ESS } & \textbf{Total}\\ 
& \textbf{cost (€)}   & \textbf{cost (€)}  & \textbf{cost (€)}& \textbf{cost (€)} & \textbf{cost (€) }    & \textbf{cost (€)}  & \textbf{cost (€)}& \textbf{cost (€)} & \textbf{cost (€)}  & \textbf{cost (€)}\\  \hline 
Case\#0&  212,184&  0.0&  16,726&  -& -&  1,791&    26,754&  -&  -& 257,456\\
Case\#1&  211,688&  0.0&  16,726&  -&  0.0&  1,791&    26,758&  -&  382& 257,344\\
Case\#2&  212,161&  0.0&  16,153&  226&  -&  1,791&  25,873&  364&  -&256,571\\
Case\#3&  211,675&  0.0&  16,158&  226&  0.0&  1,791&   25,880&  361&  369&256,464\\
 \hline 
\end{tabular}
\end{table*}

Table \ref{60bus_cases} provides the different components of the total expected cost for the different cases. In case\#0 the total cost equals $257,456$€ and wind generation curtailment occurs in both normal and post contingency states. However, thanks to the additional flexibility offered by the ESS in node 1045, the conventional operation cost reduces from 212,184€ in the Case\#0 to 211,688€ in the Case\#1. As a result, although an additional cost regarding the activation of ESS is imposed in the post contingency state, total cost decreases from 257,456€ to 257,344€. In Case\#2, although the conventional generators' cost remains almost constant, the wind generation curtailment cost is reduced meaningfully, from 26,754€ in the base case to 25,873€ in Case\#2. Consequently, the total expected cost reduces by 885€ (i.e. $257,456-256,571=885$€) as compared to the base case. In addition, in Case\#2, the wind power curtailment cost in normal operation is also reduced by 573€ (i.e. $16,726-16,153=573$€). The same trend can be observed in the last case, Case\#4, where both ESS and FL are activated in the post contingency state, where the additional volume of flexibility causes a total expected cost reduction of 992€ (i.e. from 257,456€ in Case\#0 to 256,464€ in Case\#3).

These results demonstrate that flexible resources (ESS and FL) can contribute cost-effectively to a reduction of wind energy spillage and load curtailment, improving the overall system flexibility, and allowing thereby to accommodate larger amounts of renewables. 

\begin{figure}
	\centering
	\includegraphics[width=0.90\linewidth]{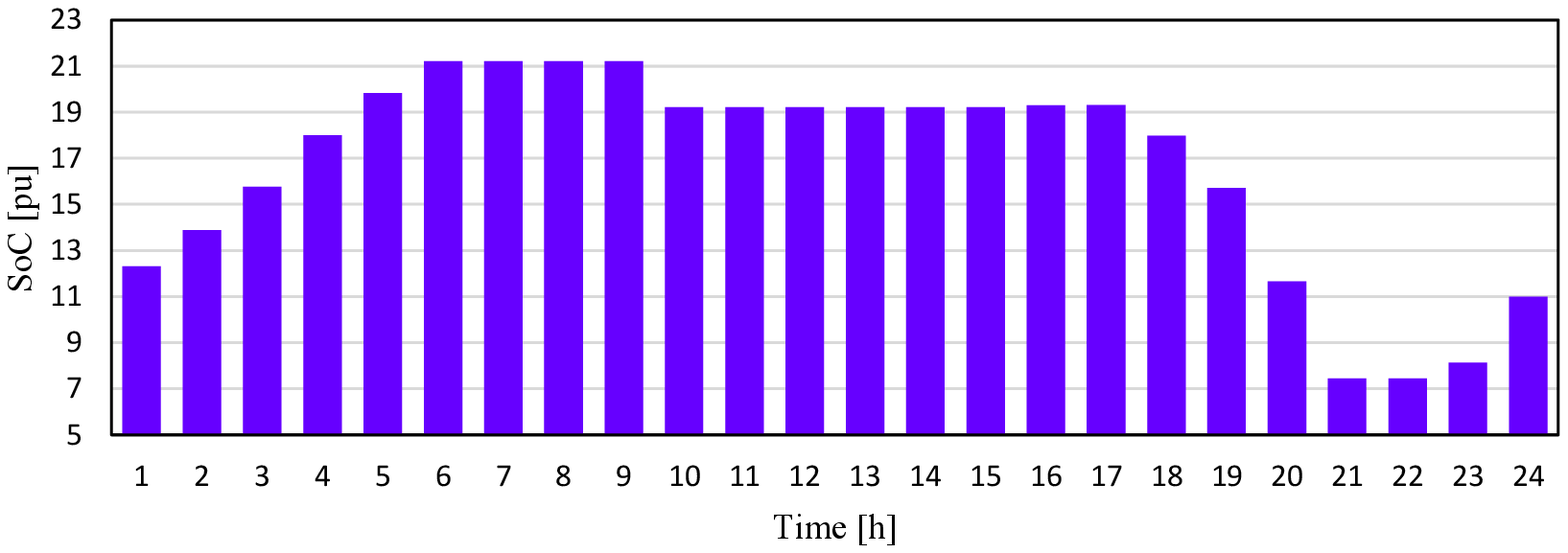}
	\caption{SOC profile of ESS at node 1045 in scenario s1 and contingency in the connecting line between nodes 1041 and 1043  }
	\label{soc_nordic}
\end{figure}

\begin{figure}
	\centering
	\includegraphics[width=0.90\linewidth]{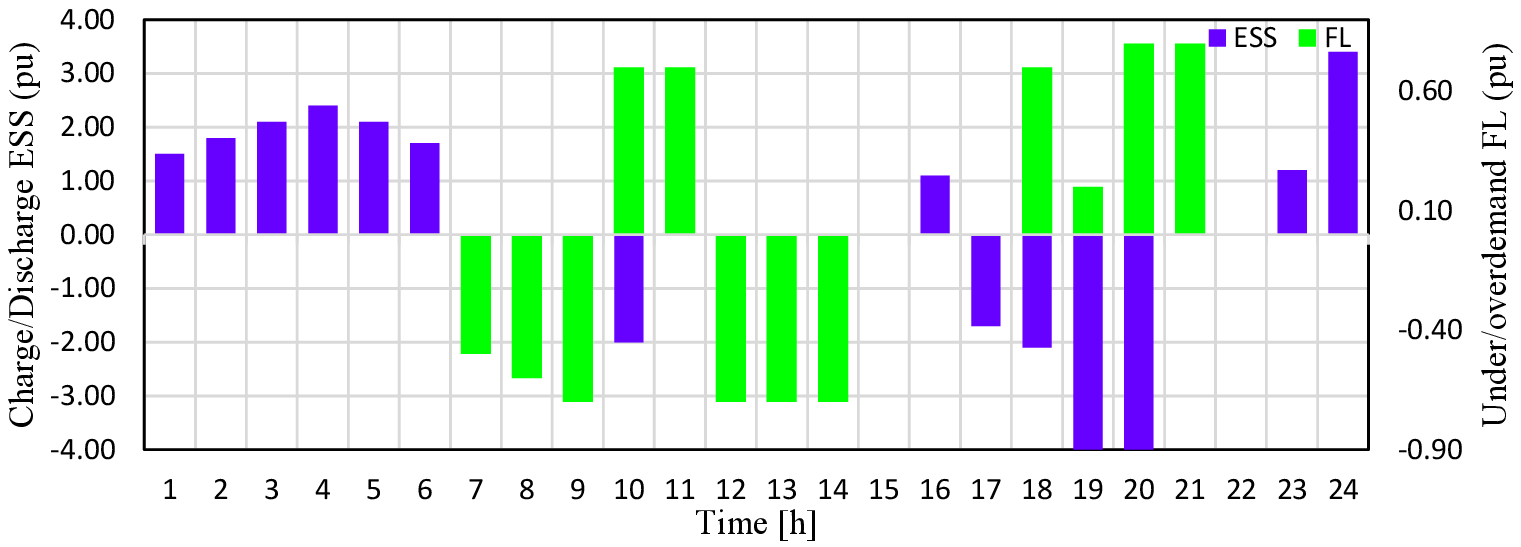}
	\caption{Injection and absorption profile of active power for both ESS and FL and contingency in the connecting line between nodes 1041 and 1043  }
	\label{ess_fl_nordic32}
\end{figure}

Figs. \ref{soc_nordic} and \ref{ess_fl_nordic32} illustrate the flexible resources behaviour at the solution of the proposed S-MP-SCOPF model. Fig. \ref{soc_nordic} shows the SoC profile of ESS at node 1045 in scenario s1 and the contingency in a single line circuit between nodes 1041 and 1043, which overloads the second line circuit in parallel between the same nodes, both colored in red in Fig. \ref{nordic32}. As expected, to alleviate the overload in the second circuit of line 1041-1043 in periods with high demand, ESS discharges during peak hours (i.e. 17-20) and charges during the lower demand hours where the line is less loaded to maintain its energy balance constraint (see Fig. \ref{soc_nordic}). 

The FL at node 1011 in the same scenario and contingency shows apparently a counter-intuitive behavior. As can be seen in Fig. \ref{ess_fl_nordic32}, FL decreases the load during hours with extra wind power generation and, to maintain its daily energy balance, increases the load in hours with high load. This behaviour can be justified as follows. In hours 7-9 and 12-14, when wind farms generate large amount of wind power, the lines between nodes 1011 and 1013, shown with red colour in Fig. \ref{nordic32}, are congested in both normal and post-contingency states. These bottlenecks require the activation of FL to remove these congestions (by creating counter-flows) and minimize the wind power curtailment. In conclusion, the primary functionality of FL is driven by the prevention of current/voltage constraints violation rather than the simpler power balance satisfaction needs.
   
\subsubsection{Model scalability} 

Table \ref{60bus_result} shows the results of the scalability test of the proposed S-MP-SCOPF model for increasing number of scenarios and thereby problem size. The results are obtained with IPOPT solver using default setting except of the relative optimality gap tolerance, which is set to $10^{-5}$. 

Note that while increasing the problem size, the elapsed time grows sharply with non-monotonic slope. For instance, although the problem size increases ten times from 1 scenario to 10 scenarios, the computation time increases more than 19.7 times (i.e. $(5,985-289)/289$).
In addition, the largest number of scenarios the solver can handle reliably is 30, which corresponds to a huge NLP optimization problem with roughly 5 millions continues variables and 9 millions of constraints, which is solved in 22,110 seconds. For larger number of scenarios the default linear (system of equations) solver package MUMPS in IPOPT fails to allocate memory even before IPOPT can start iterations. An interesting observation regarding IPOPT solver is that, as for other interior-point method-based solvers, the iteration number is little dependent on the size of the problem. 

Note that the computation time can be further significantly improved by using another linear solver in IPOPT. For example, it is reported at https://github.com/power-grid-lib/pglib-opf/blob/master/BASELINE.md that ma27 linear solver can decrease the runtime by 2-6 times as compared to default linear solver MUMPS. However, we did not manage to compile and plug ma27 linear solver in our windows code implemenation to test its performances. 

Finally, in terms of objective function one can observe that, since in cases with 1 and 2 scenarios large amount of wind power is injected into the system, generation curtailment occurs in peak hours which causes an increase in total cost for these two cases. Note that we initially generated 10 scenarios as in Table \ref{wind_scenarios}, and for the cases with larger numbers, the scenarios are replicated out of the original set of scenarios. For this reason the value of total cost remains unchanged for the cases with more than 10 scenarios. 

\begin{table}
\centering 
\caption{Nordic32 test system results for different scenarios}
\label{60bus_result}
\resizebox{\columnwidth}{!}{%
\begin{tabular}{ccccccccc} \hline
 \textbf{Number of}    & \textbf{Total  }   & \textbf{Continues}  & \textbf{Constraints }& \textbf{Iter} & \textbf{Time }   \\ 
\textbf{scenarios} & \textbf{cost (€)}  &\textbf{ variables}  &   &  &\textbf{(s)}  \\  \hline 
1&	555,874&	197,206&	287,596&	123&	289\\
2&  421,924&    394,582&    575,192&    144&764\\
3&  652,595&    591,788&    862,788&    167&1,404\\
4&  319,122&    789,164&    1,150,384&  173& 2,040\\
5&  297,430&    986,540&    1,437,980&  160&  2,283  \\
6&  284,137&    1,183,916&  1,725,576&  166&  2,878 \\
7&  274,769&    1,381,292&  2,013,172&  171&  3,596\\
8&  267,411&    1,578,668&  2,300,768&  169&  4,100\\
9&  261,006&   1,776,044&  2,588,364&  180& 4,451\\
10&  256,464&   1,973,420&  2,875,960&  186& 5,985\\
20&  256,464&   3,312,280&  5,751,920&  185& 14,351\\
30&  256,464&   4,968,420&  8,627,880&  186& 22,110\\
40&  IPOPT failed\\
 \hline 
\end{tabular}
}
\end{table}


\section{Conclusions and future work}

The research efforts devoted to address the challenge of extending the state-of-the-art in AC SCOPF (i.e. deterministic and single time period) is scarce and mostly capture one novel feature at the time. This paper has extended the state-of-the-art in AC SCOPF to capture two new dimensions (RES stochasticity and multiple time periods) as well as to model time dependent constraints of emerging sources of flexibility (FL and ESS). Accordingly, this paper solves for the first time a new NLP problem formulation in the form of stochastic multi-period AC SCOPF (S-MP-SCOPF) which we envision for procuring flexibility for ancillary services (congestion and voltage control) in renewable supply dominated power systems of the future. This problem enables computing optimal set points of the flexibility resources and other conventional control means for congestion management and voltage control in day-ahead operation planning. 

As we address a new problem, full problem details and results have been provided for a 5-node test system to foster benchmarking. The results obtained for this system show the effectiveness of the ESS and FL for flexibility provision in day-ahead operation, which are able to reduce the load curtailment cost up to 90.5\%. 

The Nordic32 test system has been used to ascertain scalability, noting that as shown in Table \ref{compare} scalability is rarely addressed by the few works that extend the state-of-the-art in AC SCOPF. The largest NLP S-MP-SCOPF problem solved (60 nodes, 34 states, 24 time periods, 30 scenarios) is roughly equivalent in size to solving an AC OPF problem for a system of huge size (cca. 1,500,000 nodes). Very few works report results for such a big NLP problem. We have relied on the state-of-the-art NLP solver IPOPT, which is widely used to solve AC OPF/SCOPF problems. The running time obtained for the largest NLP problem on this system (roughly 5 millions optimisation variables and 9 millions constraints, which is very close to the edge of computer/solver limit) is 22,110 seconds. While this time could be deemed a bit excessive for day-ahead operation planning, the elapsed time for a problem that includes three times less scenarios is 5,985 seconds, which is still acceptable. 

The paper has discussed that a massive reduction in computation time could be expected by using a more performant linear solver within IPOPT, a tailored implementation parallelizing some computations \cite{kardovs2019two}, merely using a commercial solver, or developing iterative methodologies \cite{platbrood2013generic}. 

This direct approach can thus scale to medium size systems by careful beforehand knowledge of problematic/binding contingencies, as an input from the operator, as well as reducing the number of uncertainty scenarios to a few. 

As future work, we plan to develop a tractable approach of S-MP-SCOPF problem through decomposition and approximation along the above mentioned lines. 

\bibliographystyle{ieeetr}
\bibliography{main_R4_FC}

\end{document}